\documentclass[aps,pre,twocolumn,superscriptaddress,showpacs,showkeys,amsmath,amssymb]{revtex4}

\usepackage{epsfig,amsmath,amssymb,color}
\usepackage[T1]{fontenc}
\usepackage[latin9]{inputenc}
\usepackage{epstopdf}
\renewcommand{\theequation}{\arabic{section}.\arabic{equation}}

\begin{document}

\title{Towards low-temperature peculiarities of thermodynamic quantities\\ 
       for decorated spin chains}

\author{Taras Krokhmalskii}
\affiliation{Institute for Condensed Matter Physics,
          National Academy of Sciences of Ukraine,
          Svientsitskii Street 1, 79011 L'viv, Ukraine}

\author{Taras Hutak}
\affiliation{Institute for Condensed Matter Physics,
          National Academy of Sciences of Ukraine,
          Svientsitskii Street 1, 79011 L'viv, Ukraine}

\author{Onofre Rojas}
\affiliation{Departamento de Fisica, 
          Universidade Federal de Lavras, 
          CP 3037, 37200-000, Lavras-MG, Brazil}

\author{Sergio Martins de Souza}
\affiliation{Departamento de Fisica, 
          Universidade Federal de Lavras, 
          CP 3037, 37200-000, Lavras-MG, Brazil}
          
\author{Oleg Derzhko}
\affiliation{Institute for Condensed Matter Physics,
          National Academy of Sciences of Ukraine,
          Svientsitskii Street 1, 79011 L'viv, Ukraine}
\affiliation{Department of Metal Physics, 
          Ivan Franko National University of L'viv, 
          Kyrylo \& Mephodiy Street 8, 79005 L'viv, Ukraine}

\date{\today}

\begin{abstract}
We discuss the origin of an enigmatic low-temperature behavior of one-dimensional decorated spin systems 
which was coined the pseudo-transition. 
Tracing out the decorated parts results in the standard Ising-chain model with temperature-dependent parameters
and the unexpected low-temperature behavior of thermodynamic quantities and correlations of the decorated spin chains 
can be tracked down to the critical point of the standard Ising-chain model at ${\sf H}=0$ and ${\sf T}=0$. 
We illustrate this perspective using as examples 
the spin-1/2 Ising-XYZ diamond chain,
the coupled spin-electron double-tetrahedral chain,
and
the spin-1/2 Ising-Heisenberg double-tetrahedral chain.
\end{abstract}



\maketitle

\section{Introduction}
\label{sec1}
\setcounter{equation}{0}

For a number of decorated one-dimensional spin models with short-range interactions, 
the low-temperature thermodynamic quantities exhibit an intriguing behavior 
which resembles the discontinuous or continuous temperature-driven phase transitions \cite{Galisova2015,Torrico2016,Rojas2016,Strecka2016};
it was coined the pseudo-transition \cite{Souza2018,Timonin2011}. 
Of course,
these pseudo-transitions are not the true temperature-driven transitions showing only abrupt changes or sharp maxima in thermodynamic quantities at $T=T_p>0$
as have been demonstrated in detail in several papers \cite{Souza2018,Carvalho2018,Carvalho2019,Rojas2018a,Rojas2018b,Strecka2019,Rojas2019}.
However,
a sudden increase of the entropy and the internal energy at $T=T_p$
and
an impressively fine peak of the second derivative of the free energy 
(the magnetic susceptibility and the specific heat, see, e.g., the inset in Fig.~\ref{fig09} below \cite{peaks}) 
at $T=T_p$ 
are rather striking features which call for explanations \cite{Souza2018}.
Further on, it was found 
that the correlation functions at $T=T_p$ have very large correlation length, 
i.e., decrease very slowly with the distance increase \cite{Carvalho2019},
and
that the zero-temperature phase boundary residual entropy may serve as an indicator of the pseudo-transition \cite{Rojas2018a}.
Moreover,
in the vicinity of (but not too close to) $T_p$,
both ascending as well as descending part of the peaks fits precisely a power-law behavior.
Calculations for four specific models
(Ising-XYZ diamond chain, coupled spin-electron double-tetrahedral chain, Ising-XXZ two-leg ladder, and Ising-XXZ three-leg tube)
yield a universal set of pseudo-critical exponents
with the values
$\alpha=\alpha^\prime=3$ for the specific heat,
$\gamma=\gamma^\prime=3$ for the susceptibility, 
and
$\nu=\nu^\prime=1$ for the correlation length \cite{Rojas2018b}.
It is worth noting here that decorated Ising-chain models may be realized in certain real magnetic compounds containing lanthanide ions \cite{Heuvel2010}.

What remains outside of those studies, in our opinion, is the reasons for the emergence of pseudo-transitions.
With the present study we wish to fill in this gap 
illustrating what is behind the enigmatic low-temperature dependences of the one-dimensional systems exhibiting pseudo-transition.
An important step in our consideration is a mapping of the decorated spin chains onto an Ising-chain model
(Section~\ref{sec2} and three Appendices~A, B, and C).
Although this mapping was mentioned in all previous studies,
however, to our mind, it was not enough appreciated.
The distinctive feature of the resulting effective Ising-chain model is the temperature-dependent parameters.
Pseudo-transitions are observed when the effective exchange is ferromagnetic and the effective field changes its sign at certain temperature,
Eq.~(\ref{202}) (Section~\ref{sec2}). 
If this temperature is low enough,
we face remnants of the critical point of the standard Ising-chain model.
Moreover,
the temperature-dependent parameters lead to interesting relations 
between the internal energy, the entropy, and the specific heat on one side 
and 
the magnetization and the susceptibility on the other side,
Eqs.~(\ref{305}), (\ref{306}), (\ref{307}) (Section~\ref{sec3}).
These relations provide the background for understanding universality found in Ref.~\cite{Rojas2018b} (Section~\ref{sec4}).
The elaborated perspective 
unveils the ``mystery'' of pseudo-transitions 
and 
yields a useful tool for revealing new decorated-spin-model candidates with peculiar low-temperature behavior  
to be explored theoretically and, hopefully, experimentally (Section~\ref{sec5}).

\section{Effective Ising-chain model}
\label{sec2}
\setcounter{equation}{0}

Decorated spin models in the regime when they exhibit temperature-driven pseudo-transitions
(for example, 
the spin-1/2 Ising-XYZ diamond chain, see Appendix~A,
the coupled spin-electron double-tetrahedral chain, see Appendix~B,
or
the spin-1/2 Ising-Heisenberg double-tetrahedral chain, see Appendix~C)
can be rigorously reduced to the effective Ising-chain model with the Hamiltonian
\begin{eqnarray}
\label{201}
{\cal H}_{\rm eff}
=C- J_{\rm eff}\sum_{n}\sigma_n\sigma_{n+1}- H_{\rm eff}\sum_n\sigma_n,
\end{eqnarray}
where
$n=1,\ldots,{\sf N}$,
$C=C(T)$,
$J_{\rm eff}=J_{\rm eff}(T)>0$ is ferromagnetic, 
whereas
$H_{\rm eff}=H_{\rm eff}(T)$ does change its sign while the temperature $T$ grows.
Then the pseudo-critical temperature $T_p$ is defined by \cite{Souza2018,Carvalho2019}
\begin{eqnarray}
\label{202}
H_{\rm eff}(T_p)=0.
\end{eqnarray}
Equation~(\ref{202}) provides the necessary condition for occurrence of the pseudo-transition.
Moreover, $J_{\rm eff}(T)$ varies slowly and does not change its sign.
Temperature dependences of the constant term, the ferromagnetic exchange, and the magnetic field reflect a certain internal structure of the initial model 
(i.e., the decorated spin chain),
which is hidden now in the specific functions $C(T)$, $J_{\rm eff}(T)$, and $H_{\rm eff}(T)$.

Using the vocabulary
\begin{eqnarray}
\label{203}
\frac{J_{\rm eff}(T)}{T}=\frac{\sf J}{\sf T},
\;\;\;
\frac{H_{\rm eff}(T)}{J_{\rm eff}(T)}=\frac{\sf H}{\sf J},
\end{eqnarray}
we may introduce the standard ferromagnetic Ising-chain model,
\begin{eqnarray}
\label{204}
{\cal H}({\sf N},{\sf J},{\sf H})=-{\sf J}\sum_{n}\sigma_n\sigma_{n+1}-{\sf H}\sum_n\sigma_n,
\;\;\;
{\sf J}>0,
\end{eqnarray}
$n=1,\ldots,{\sf N}$,
periodic boundary conditions are implied,
$\sigma_n=\pm 1$,
which is explained in most textbooks on statistical mechanics \cite{Baxter1982}.
The model is exactly solvable by the transfer-matrix method.
Knowing the eigenvalues of the transfer matrix
\begin{eqnarray}
\label{205}
\lambda_{\pm}
=
\exp\frac{{\sf J}}{{\sf T}} \left[\cosh\frac{{\sf H}}{{\sf T}} \pm\sqrt{\sinh^2\frac{{\sf H}}{{\sf T}}+\exp\left(-\frac{4{\sf J}}{{\sf T}}\right)}\right],
\end{eqnarray}
one immediately gets all required quantities,
e.g., 
the Helmholtz free energy
${\sf F}({\sf T},{\sf H},{\sf N})/{\sf N} \to -{\sf T} \ln\lambda_+$
or the pair spin correlations at the distance $m$
which behaves as $\propto(\lambda_-/\lambda_+)^m$.
Around the critical point, ${\sf H}={\sf T}=0$,
the behavior of thermodynamic quantities is characterized by the set of critical exponents:
$\alpha=1$, $\beta=0$, $\gamma=1$, $\nu=1$, and $\eta=1$ \cite{Baxter1982}.

Usually, only the region ${\sf H}\ge 0$ is discussed, since the results for ${\sf H}\le 0$ follow directly by symmetry arguments.
However, for the case at hand it would be convenient to consider further both signs of ${\sf H}$ explicitly.

\begin{figure}
\begin{center}
\includegraphics[clip=true,width=0.95\columnwidth]{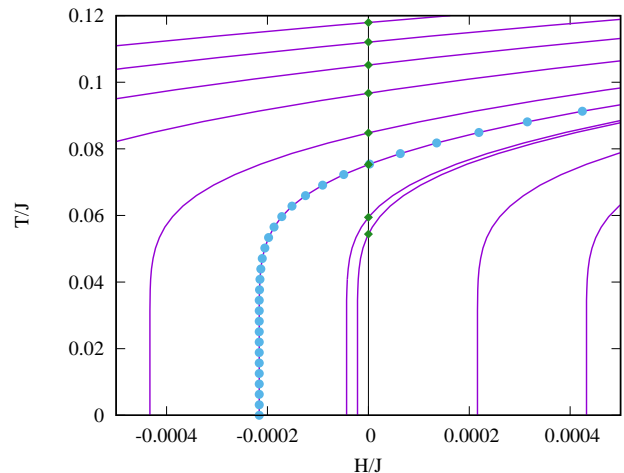}
\caption{The spin-1/2 Ising-XYZ diamond chain described in Appendix~A, 
which is represented by Eqs.~(\ref{201}), (\ref{a02}), and (\ref{a03}), 
in the ${\sf H}/{\sf J}-{\sf T}/{\sf J}$ plane.
We consider the following set of parameters for the initial model: $J=100$, $J_z=24$, $J_0=-24$, $\gamma=0.7$. 
Violet lines (trajectories) correspond to the following values of $h$:
12.7, 12.71, 12.72, 12.73, 12.74, 12.745, 12.749, 12.749\,5, 12.755, 12.76.
Sky blue circles correspond to $T=0, 0.01, 0.02,\ldots,0.29$ (for $h=12.745$).
Green diamonds correspond to pseudo-critical temperatures $T_p$, see Eq.~(\ref{202}), for different $h<12.75$.}
\label{fig01}
\end{center}
\end{figure}

\begin{figure}
\begin{center}
\includegraphics[clip=true,width=0.95\columnwidth]{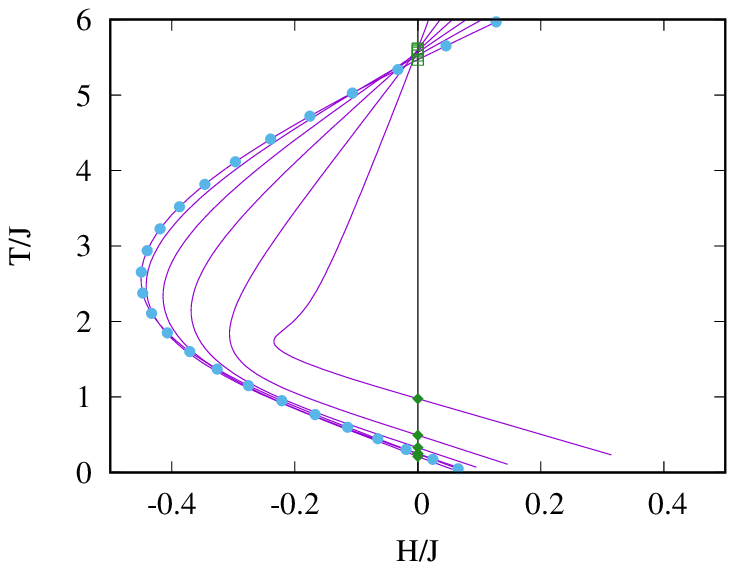}\\
\vspace{1mm}
\includegraphics[clip=true,width=0.95\columnwidth]{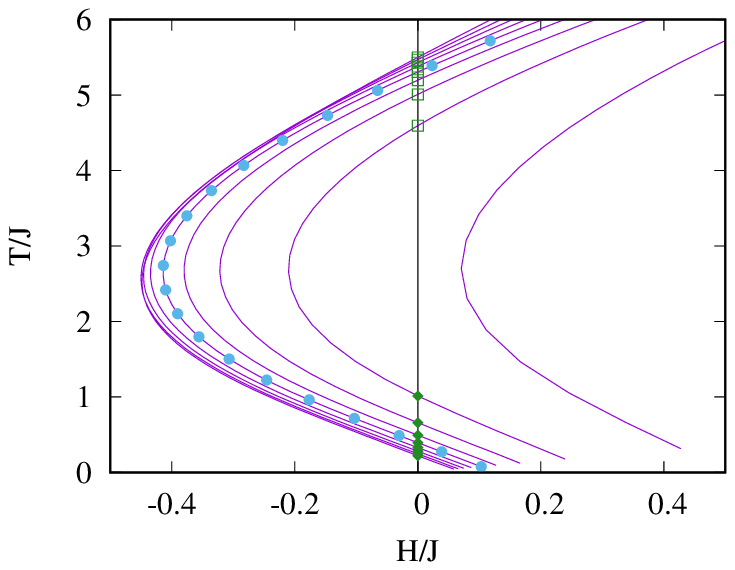}
\caption{The double-tetrahedral chain of localized Ising spins and mobile electrons described in Appendix~B,
which is represented by Eqs.~(\ref{201}), (\ref{b02}), and (\ref{b03}), 
in the ${\sf H}/{\sf J}-{\sf T}/{\sf J}$ plane.
We consider the following set of parameters for the initial model: $J=1$, $t=0.6$, $U=5$.
Violet lines (trajectories) correspond to the following values of $H$:
0.2, 0.4, 0.6, 0.8, 1, 1.2 (top)
and
1.1, 1.2, 1.3, 1.4, 1.5, 1.6, 1.7, 1.8, 1.9 (bottom).
Sky blue circles correspond to $T=0.01,\,0.035,\,0.06,\,\ldots,0.285$ 
(for $H=1$ in the top panel and for $H=1.5$ in the bottom panel).
Green diamonds and squares correspond to pseudo-critical temperatures $T_p$, see Eq.~(\ref{202}), for different $H<1.884$.}
\label{fig02}
\end{center}
\end{figure}

\begin{figure}
\begin{center}
\includegraphics[clip=true,width=0.95\columnwidth]{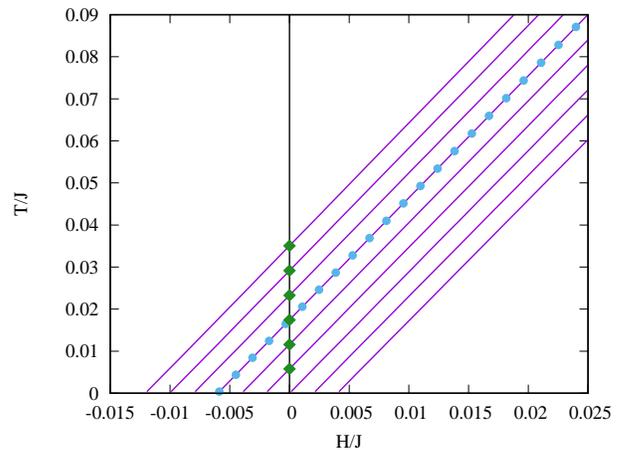}
\caption{The spin-1/2 Ising-Heisenberg double-tetrahedral chain described in Appendix~C, 
which is represented by Eqs.~(\ref{201}), (\ref{c03}), and (\ref{c02}),
in the ${\sf H}/{\sf J}-{\sf T}/{\sf J}$ plane. 
We consider the following set of parameters for the initial model: $h=h_{z}=20$, $J_{0}=J=-10$.
Violet lines (trajectories) correspond to the following values of $J_{z}$: 
$-14.94$, $-14.95$, $-14.96$, $-14.97$, $-14.98$, $-14.99$, $-15.00$, $-15.01$, $-15.02$. 
Sky blue circles correspond to $T=0, 0.01, 0.02,\dots,0.21$ (for $J_z=-14.97$). 
Green diamonds correspond to pseudo-critical temperature $T_{p}$, see Eq.~(\ref{202}), for different $J_z>-15.00$.}
\label{fig03}
\end{center}
\end{figure}

Using the relations given in Eq.~(\ref{203}),
we can construct the trajectories in the ${\sf H}/{\sf J}-{\sf T}/{\sf J}$ plane,
along which the initial system 
[or, equivalently, the effective system (\ref{201}) with $T$-dependent parameters]
moves as $T$ grows from low to high values passing $T_p$.
Some of such trajectories 
for the spin-1/2 Ising-XYZ diamond chain (see Appendix~A),
for the coupled spin-electron double-tetrahedral chain (see Appendix~B),
and
for the spin-1/2 Ising-Heisenberg double-tetrahedral chain (see Appendix~C)
are shown in Fig.~\ref{fig01}, in Fig.~\ref{fig02}, and in Fig.~\ref{fig03}, respectively.
In what follows,
we use the ${\sf H}/{\sf J}-{\sf T}/{\sf J}$ plane to demonstrate certain $T$-dependences for the effective model (and thus for the initial model)
moving along such trajectories
(see, e.g., Figs.~\ref{fig04}, \ref{fig05}, \ref{fig06} below
which regard to the case of the spin-1/2 Ising-XYZ diamond chain).

It is worth making few remarks here.
First of all,
the reported trajectories in the ${\sf H}/{\sf J}-{\sf T}/{\sf J}$ plane permit one to compare different decorated models.
For instance,
comparing Figs.~\ref{fig01} and \ref{fig02},
one notes several important differences.
Equation~(\ref{202}) for the double-tetrahedral chain case has two solutions 
(green filled diamonds and empty squares in Fig.~\ref{fig02}),
although the higher-temperature one (green empty squares) does not manifest itself in the observed properties of the decorated spin chain,
see Eq.~(\ref{310}) below. 
The temperature scale for two models is obviously different 
which results in ``stronger'' peculiarities for the spin-1/2 Ising-XYZ diamond chain
since they occur at lower temperatures ${\sf T}/{\sf J}$.
Moreover,
while for the first model ${\sf T}/{\sf J}$ which corresponds to $T_p$ can be made infinitesimally small (Fig.~\ref{fig01}),
for the second model ${\sf T}/{\sf J}$ cannot be lower than ${\sf T}/{\sf J}=0.211\ldots$ (Fig.~\ref{fig02}).
As it will be seen later, 
a slope of the trajectory at the point where it crosses the straight vertical line ${\sf H}=0$ may be also important.
While for the spin-1/2 Ising-XYZ diamond chain case the slope obviously decreases as $h$ tends to $h=12.75$,
for the double-tetrahedral chain case the slope is less sensitive to the value of $H<1.884$.

\section{Effective Ising-chain model around $T_p$}
\label{sec3}
\setcounter{equation}{0}

The properties of the effective Ising-chain model (\ref{201})
(and thus of the initial decorated model) 
are determined by the eigenvalues of the transfer matrix $\lambda_{\pm}$ (\ref{205}), (\ref{203}).
They straightforwardly yield 
the Helmholtz free energy per site
\begin{eqnarray}
\label{301}
f=\frac{C}{\sf N}-T\ln\lambda_+
\end{eqnarray}
[here the first term has appeared because of the constant term $C$ in Eq.~\eqref{201}]
or
the correlation length
\begin{eqnarray}
\label{302}
\xi=\frac{1}{\ln\frac{\lambda_+}{\lambda_-}}.
\end{eqnarray}
Taking the derivatives with respect to the field, 
we immediately obtain the Ising-spin magnetization and susceptibility
\begin{eqnarray}
\label{303}
m=-\frac{\partial f}{\partial H_{\rm eff}}
=\frac{\sinh \frac{H_{\rm eff}}{T}}{\sqrt{\sinh^2\frac{H_{\rm eff}}{T}+\exp\left(-\frac{4J_{\rm eff}}{T}\right)}},
\end{eqnarray}
\begin{eqnarray}
\label{304}
\chi=\frac{\partial m}{\partial H_{\rm eff}}
=\frac{1}{T}
\frac{\cosh\frac{H_{\rm eff}}{T}\exp\left(-\frac{4J_{\rm eff}}{T}\right)}
{\left[\sinh^2\frac{H_{\rm eff}}{T}+\exp\left(-\frac{4J_{\rm eff}}{T}\right)\right]^{\frac{3}{2}}},
\end{eqnarray}
see Ref.~\cite{Baxter1982}.

The only important peculiarity of the effective Ising-chain model \eqref{201} is related to the temperature dependences of the effective parameters.
Therefore one has to take the derivatives with respect to the temperature with cautious.
Thus, the internal energy, the entropy, and the specific heat are given by the following formulas:
\begin{eqnarray}
\label{305}
e
=
-T^2\frac{\partial}{\partial T}\frac{f}{T}
-T\frac{\partial f}{\partial J_{\rm eff}}\frac{\partial J_{\rm eff}}{\partial T}
-T\frac{\partial f}{\partial H_{\rm eff}}\frac{\partial H_{\rm eff}}{\partial T}
\nonumber\\
=
-T^2\frac{\partial}{\partial T}\frac{f}{T} - T\frac{\partial f}{\partial J_{\rm eff}} J_{\rm eff}^{\prime} + Tm H_{\rm eff}^{\prime}
\nonumber\\
=\sum_{j=1}^3e^{(j)}(T),
\end{eqnarray}
\begin{eqnarray}
\label{306}
s=-\frac{\partial f}{\partial T} - \frac{\partial f}{\partial J_{\rm eff}} J_{\rm eff}^{\prime} + m H_{\rm eff}^{\prime}
\nonumber\\
=\sum_{j=1}^3s^{(j)}(T),
\end{eqnarray}
\begin{eqnarray}
\label{307}
c=-T\frac{\partial^2 f}{\partial T^2}
\nonumber\\
-2T\frac{\partial^2 f}{\partial J_{\rm eff}\partial T}J_{\rm eff}^{\prime} +2T\frac{\partial m}{\partial T}H_{\rm eff}^{\prime}
-T\frac{\partial f}{\partial J_{\rm eff}}J_{\rm eff}^{\prime\prime}
+TmH_{\rm eff}^{\prime\prime}
\nonumber\\
-T\frac{\partial^2 f}{\partial J_{\rm eff}^2}(J_{\rm eff}^{\prime})^2
+2T\frac{\partial m}{\partial J_{\rm eff}}J_{\rm eff}^{\prime}H_{\rm eff}^{\prime}
+T\chi(H_{\rm eff}^\prime)^2
\nonumber\\
=\sum_{j=1}^8c^{(j)}(T).
\end{eqnarray}
Interestingly, 
according to Eqs.~(\ref{305}), (\ref{306}), and (\ref{307}),
the internal energy and the entropy are related to the magnetization
[the terms $e^{(3)}(T)$ and $s^{(3)}(T)$],
whereas 
the specific heat is related to the susceptibility
[the term $c^{(8)}(T)$]; 
obviously, this happens owing to the temperature-dependent Hamiltonian parameters.

Now we can discuss the temperature dependences of various quantities for the decorated spin chains which can be presented as the effective model (\ref{201}). 
Consider first the correlation length $\xi$,
\begin{eqnarray}
\label{308}
\frac{1}{\xi}
=\ln
\frac{\cosh\frac{H_{\rm eff}}{T}+\sqrt{\sinh^2\frac{H_{\rm eff}}{T}+\exp\left(-\frac{4J_{\rm eff}}{T}\right)}}
{\cosh\frac{H_{\rm eff}}{T}-\sqrt{\sinh^2\frac{H_{\rm eff}}{T}+\exp\left(-\frac{4J_{\rm eff}}{T}\right)}}.
\end{eqnarray}
At the pseudo-critical temperature $T_p$ (\ref{202}), 
Eq.~(\ref{308}) becomes
\begin{eqnarray}
\label{309}
\frac{1}{\xi(T_p)}
=\ln
\frac{1+\exp\left(-\frac{2J_{\rm eff}(T_p)}{T_p}\right)}{1-\exp\left(-\frac{2J_{\rm eff}(T_p)}{T_p}\right)}.
\end{eqnarray}
Evidently,
$\xi(T_p)$ tends to infinity only for $2J_{\rm eff}(T_p)/T_p\to \infty$.
We may suggest as a sufficient condition for the pseudo-transition the following one:
\begin{eqnarray}
\label{310}
\frac{2J_{\rm eff}(T_p)}{T_p} \gg 1,
\end{eqnarray}
where $T_p$ is defined in Eq.~(\ref{202}).
Obviously,
Eq.~(\ref{310}) says that ${\sf T}/{\sf J}\ll 2$ at $T=T_p$,
i.e., that the temperature ${\sf T}$ which corresponds to $T_p$ is sufficiently low.
Under this assumption, Eq.~(\ref{309}) gives the following estimate for $\xi(T_p)$
\begin{eqnarray}
\label{311}
\xi(T_p)\approx \frac{1}{2} \exp\frac{2J_{\rm eff}(T_p)}{T_p}<\infty.
\end{eqnarray}
However, in the vicinity of $T_p$, when 
\begin{eqnarray}
\label{312}
\sinh^2\frac{H_{\rm eff}}{T}\gg \exp\left(-\frac{4J_{\rm eff}}{T}\right),
\end{eqnarray}
Eq.~\eqref{308} reads
\begin{eqnarray}
\label{313}
\frac{1}{\xi}
\approx\ln
\frac{\cosh\frac{H_{\rm eff}}{T}+\left\vert\sinh\frac{H_{\rm eff}}{T}\right\vert}{\cosh\frac{H_{\rm eff}}{T}-\left\vert\sinh\frac{H_{\rm eff}}{T}\right\vert}
\approx\frac{2\vert H_{\rm eff}\vert}{T}
\end{eqnarray}
resulting in
\begin{eqnarray}
\label{314}
\xi\vert_{T\approx T_p, (\ref{312})}\approx\frac{T}{2\vert H_{\rm eff}\vert}\propto\frac{1}{\vert H_{\rm eff}(T)\vert}.
\end{eqnarray}
Clearly, this quantity is large while $T$ approaches condition (\ref{202}),
however, precisely at $T=T_p$, the inequality (\ref{312}) fails and Eq.~(\ref{309}) holds implying finite correlation length $\xi(T_p)$.

\begin{figure}
\begin{center}
\includegraphics[clip=true,width=0.95\columnwidth]{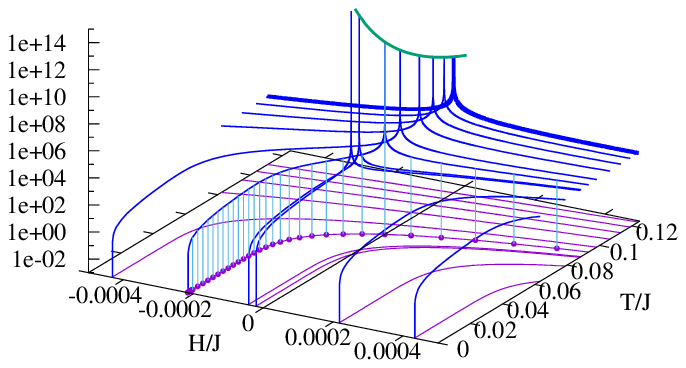}\\
\vspace{1mm}
\includegraphics[clip=true,width=0.95\columnwidth]{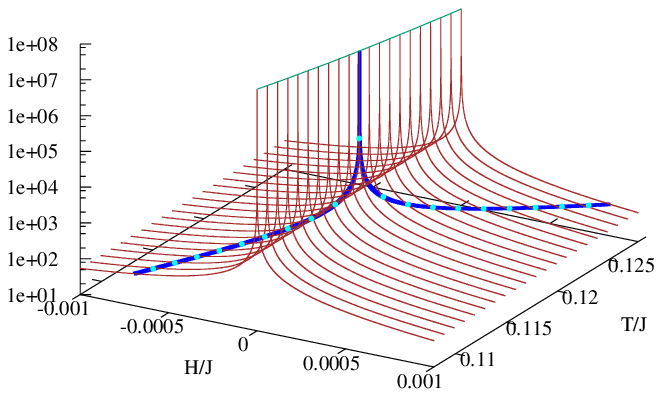}
\caption{Correlation length $\xi$ in the ${\sf H}/{\sf J}-{\sf T}/{\sf J}$ plane; 
the spin-1/2 Ising-XYZ diamond chain (with $\gamma=0.7$) is described in Appendix~A, see Fig.~\ref{fig13}.
Ten violet curves in the upper panel correspond to the following values of $h$: 
12.7, 12.71, 12.72, 12.73, 12.74, 12.745, 12.749, 12.749\,5, 12.755, 12.76.
The violet curve corresponding to $h=12.745$ in the upper panel has also violet circles,
which correspond to the monotonical increase of $T$: 0, 0.01, 0.02, \ldots, 0.29,
cf. Fig.~\ref{fig01}.
Blue curves present dependences $\xi(T)$ at fixed $h$ given above.
The thick blue curve in the upper panel presents $\xi(T)$ at $h=12.7$.
Green curve in the both panels is the dependence $\xi({\sf T}/{\sf J})$ at ${\sf H}=0$ for the standard Ising-chain model.
In the lower panel we show the dependence $\xi(T)$ at $h=12.7$ in detail using another scale.
Moreover, by brown curves we show the dependence $\xi({\sf H})$ for the standard Ising-chain model at several ${\sf T}$: 
0.107\,9, 0.108\,9, \ldots, 0.127\,9.
Intersections of the blue curve with the set of brown curves are denoted by cyan circles.}
\label{fig04}
\end{center}
\end{figure}

In Fig.~\ref{fig04},
we show the dependence $\xi(T)$ 
for the spin-1/2 Ising-XYZ diamond chain for a representative set of parameters when the model shows pseudo-transition (see Fig.~\ref{fig13})
by blue curves ($\xi$, ordinates) upon violet curves ($T$, abscissas)
in the ${\sf H}/{\sf J}-{\sf T}/{\sf J}$ plane.
$\xi$ increases as $T$ approaches $T_p$ and reaches its maximal value (\ref{309}) at $T=T_p$.
Moreover, 
in the lower panel we also show the dependence $\xi({\sf H})$ for the standard Ising-chain model (\ref{204}) 
at few values of ${\sf T}/{\sf J}=0.107\,9, 0.108\,9, \ldots, 0.127\,9$ (brown curves)
to reveal the relation between the models (\ref{201}) (thick blue curve) and (\ref{204}) (brown curves).
For example, for $h=12.7$, $T_p\approx 0.372\,6$ that correspond to ${\sf H}=0$ and ${\sf T}/{\sf J}\approx 0.117\,9$.
As it is clear from Fig.~\ref{fig04},
the Ising-chain model singularity at ${\sf H}=0$ and ${\sf T}=0$ indicated by the green curves $\xi({\sf T}/{\sf J})$ at ${\sf H}=0$
is related to an abrupt increase of $\xi(T)$ for the effective model (\ref{201}) at $T_p$.
However,
one should not expect for the effective model (\ref{201}) 
the critical behavior inherent in the standard Ising-chain model, see Sec.~\ref{sec4}.

\begin{figure}
\begin{center}
\includegraphics[clip=true,width=0.95\columnwidth]{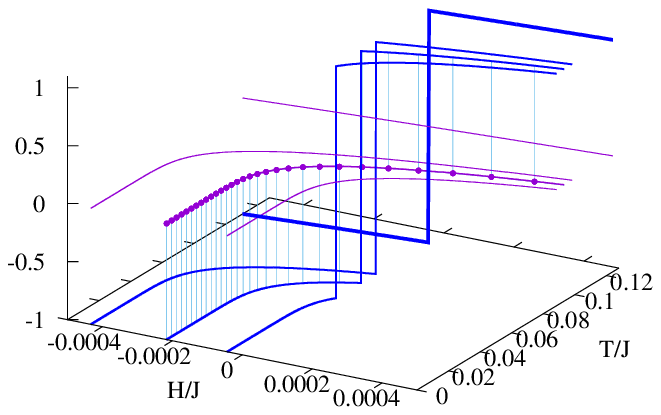}\\
\vspace{1mm}
\includegraphics[clip=true,width=0.95\columnwidth]{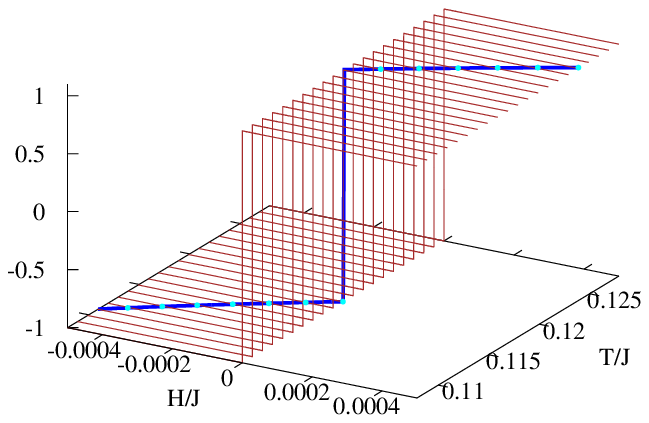}
\caption{Magnetization $m$ in the ${\sf H}/{\sf J}-{\sf T}/{\sf J}$ plane; 
the spin-1/2 Ising-XYZ diamond chain (with $\gamma=0.7$) is described in Appendix~A, see Fig.~\ref{fig13}.
Four violet curves in the upper panel correspond to the following values of $h$: 
12.7, 12.74, 12.745, 12.749.
The violet curve corresponding to $h=12.745$ in the upper panel has also violet circles,
which correspond to the monotonical increase of $T$: 0, 0.01, 0.02, \ldots, 0.29, 
cf. Fig.~\ref{fig01}.
Blue curves present dependences $m(T)$ at fixed $h$ given above.
The thick blue curve in the upper panel presents $m(T)$ at $h=12.7$.
In the lower panel we show the dependence $m(T)$ at $h=12.7$ in detail using another scale.
Moreover, by brown curves we show the dependence $m({\sf H})$ for the standard Ising-chain model at several ${\sf T}$:
0.107\,9, 0.108\,9, \ldots, 0.127\,9.
Intersections of the blue curve with the set of brown curves are denoted by cyan circles.}
\label{fig05}
\end{center}
\end{figure}

The magnetization of the initial model is given by the magnetization of the effective model (\ref{201})
[the Ising-spin magnetization of the model (\ref{a01}) is two times smaller than the magnetization of the effective model (\ref{201})].
In Fig.~\ref{fig05},
we show the temperature dependence of the magnetization of the initial model $m(T)$ at $h=12.7,\,12.74,\,12.745,\,12.749$.
With temperature grow,
$H_{\rm eff}$ changes its sign at $T_p$ resulting in a well-pronounced jump of the magnetization from almost $-1$ to almost $1$
(since the values of $\sf T$ which correspond $T_p$ are rather small;
e.g., the initial model with $h=12.7$ exhibits the jump at $T_p\approx 0.372\,6$ which corresponds to ${\sf T}/{\sf J}\approx 0.117\,9$,
see the lower panel in Fig.~\ref{fig05}).
The well-pronounced jump in the temperature dependence of magnetization is simply because of the change of the sign of the field at rather low temperatures.
On the other hand,
it has important consequences for the temperature dependence of the internal energy and the entropy,
see the third terms $e^{(3)}(T)$ and $s^{(3)}(T)$ in Eqs.~(\ref{305}) and (\ref{306}).  

\begin{figure}
\begin{center}
\includegraphics[clip=true,width=0.95\columnwidth]{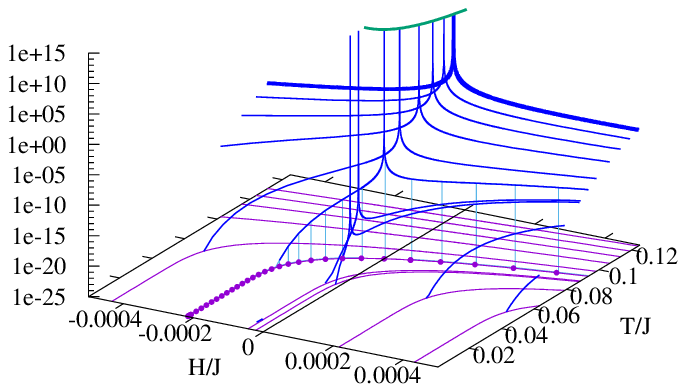}\\
\vspace{1mm}
\includegraphics[clip=true,width=0.95\columnwidth]{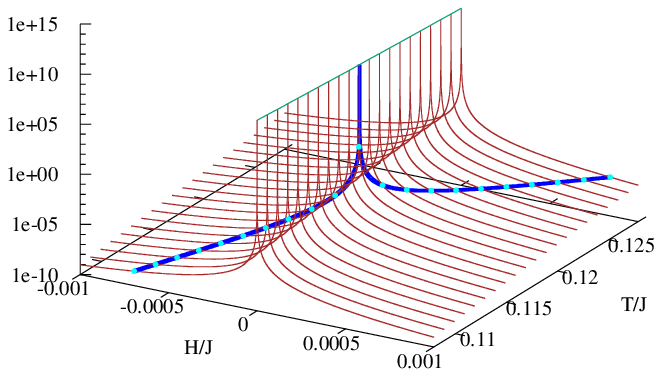}
\caption{Susceptibility $\chi$ in the ${\sf H}/{\sf J}-{\sf T}/{\sf J}$ plane; 
the spin-1/2 Ising-XYZ diamond chain (with $\gamma=0.7$) is described in Appendix~A, see Fig.~\ref{fig13}.
Ten violet curves in the upper panel correspond to the following values of $h$: 
12.7, 12.71, 12.72, 12.73, 12.74, 12.745, 12.749, 12.749\,5, 12.755, 12.76.
The violet curve corresponding to $h=12.745$ in the upper panel has also violet circles,
which correspond to the monotonical increase of $T$: 0, 0.01, 0.02, \ldots, 0.29,
cf. Fig.~\ref{fig01}.
Blue curves present dependences $\chi(T)$ at fixed $h$ given above.
The thick blue curve in the upper panel presents $\chi(T)$ at $h=12.7$.
Green curve in the both panels is the dependence $\chi({\sf T}/{\sf J})$ at ${\sf H}=0$ for the standard Ising-chain model.
In the lower panel we show the dependence $\chi(T)$ at $h=12.7$ in detail using another scale.
Moreover, by brown curves we show the dependence $\chi({\sf H})$ for the standard Ising-chain model at several ${\sf T}$:
0.107\,9, 0.108\,9, \ldots, 0.127\,9.
Intersections of the blue curve with the set of brown curves are denoted by cyan circles.}
\label{fig06}
\end{center}
\end{figure}

We pass to the susceptibility (\ref{304})
[the Ising-spin susceptibility of the model (\ref{a01}) is four times smaller than the susceptibility of the effective model (\ref{201})].
At $T=T_p$, we have
\begin{eqnarray}
\label{315}
\chi(T_p)=\frac{1}{T_p}\exp\frac{2J_{\rm eff}(T_p)}{T_p}\approx \frac{2\xi(T_p)}{T_p}.
\end{eqnarray}
In the vicinity of $T_p$ when Eq.~(\ref{312}) holds,
we have
\begin{eqnarray}
\label{316}
\chi\vert_{T\approx T_p, (\ref{312})}
\approx
\frac{1}{T}
\frac{\cosh\frac{H_{\rm eff}}{T}}{\vert\sinh\frac{H_{\rm eff}}{T}\vert^3} \exp\left(-\frac{4J_{\rm eff}}{T}\right)
\nonumber\\
\approx
\frac{1}{T}
\left\vert\frac{T}{H_{\rm eff}}\right\vert^3  \exp\left(-\frac{4J_{\rm eff}}{T}\right)
\propto
\frac{1}{\vert H_{\rm eff}(T)\vert^3}.
\end{eqnarray}
Again,
this quantity is large as $T$ approaches $T_p$,
however, at $T_p$ we have the finite value given in Eq.~(\ref{315}).

Interestingly, the approximate results in Eqs.~(\ref{313}) and (\ref{316}) are valid in a wider range of temperatures $T$ 
(not only in the vicinity of $T_p$)
if one uses here $H_{\rm eff}$ and $J_{\rm eff}$ given by Eqs.~(\ref{a02}) and (\ref{a03}).

The temperature dependencies of the susceptibility of the initial model at various $h=12.7,\ldots,12.76$ are shown in Fig.~\ref{fig06}.
As temperature $T$ grows approaching $T_p$,
$\chi(T)$ abruptly increases at $T_p$ 
achieving the value of $\chi$ for the standard Ising-chain model at ${\sf H}=0$ and rather low temperature ${\sf T}/{\sf J}=T_p/J_{\rm eff}(T_p)$
(green curves in Fig.~\ref{fig06}).
However, this abrupt increase, although related to the criticality at ${\sf H}=0$ and ${\sf T}=0$, is by no means identical to it.
This is nicely seen, e.g., in the lower panel of Fig.~\ref{fig06},
compare the blue curve $\chi(T)$ with the green curve $\chi({\sf T})$ at ${\sf H}=0$.
We emphasize in passing
that the large values of $\chi(T_p)$ may manifest themselves in the temperature dependence of the specific heat,
see the eighth term $c^{(8)}(T)$ in Eq.~(\ref{307}).

\begin{figure}
\begin{center}
\includegraphics[clip=true,width=0.95\columnwidth]{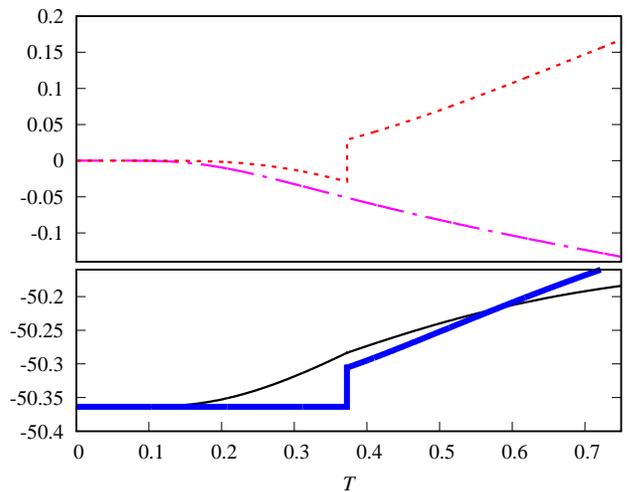}
\caption{Internal energy $e$ versus $T$; 
the spin-1/2 Ising-XYZ diamond chain (with $\gamma=0.7$, $h=12.7$) is described in Appendix~A, see Fig.~\ref{fig13}.
Thick blue curve corresponds to $e(T)$,
thin black, magenta, and red curves correspond to $e^{(1)}(T)$, $e^{(2)}(T)$, and $e^{(3)}(T)$, respectively,
see Eq.~(\ref{305}).}
\label{fig07}
\end{center}
\end{figure}

\begin{figure}
\begin{center}
\includegraphics[clip=true,width=0.95\columnwidth]{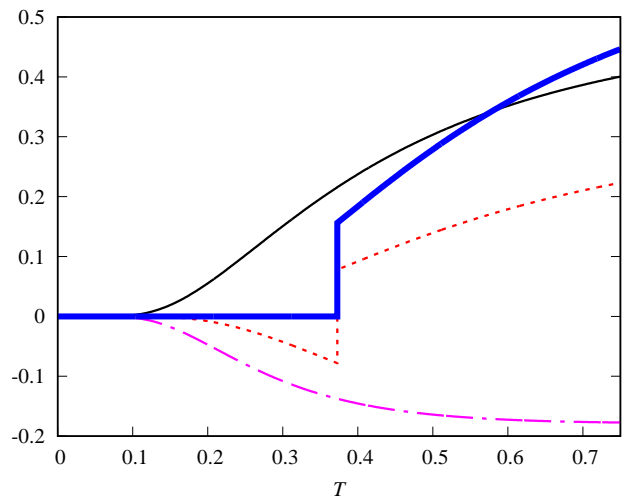}
\caption{Entropy $s$ versus $T$; 
the spin-1/2 Ising-XYZ diamond chain (with $\gamma=0.7$, $h=12.7$) is described in Appendix~A, see Fig.~\ref{fig13}.
Thick blue curve corresponds to $s(T)$,
thin black, magenta, and red curves correspond to $s^{(1)}(T)$, $s^{(2)}(T)$, and $s^{(3)}(T)$, respectively,
see Eq.~(\ref{306}).}
\label{fig08}
\end{center}
\end{figure}

The temperature dependence of the internal energy and the entropy can be easily understood 
on the basis of Eqs.~(\ref{305}) and (\ref{306}) complemented by Fig.~\ref{fig05}.
The jump in the temperature dependence of $m$ at $T=T_p$ immediately generates the jump in the temperature dependence of $e$ and $s$.
In Figs.~\ref{fig07} and \ref{fig08}, 
we show all three contributions $e^{(1)}(T)$, $e^{(2)}(T)$, $e^{(3)}(T)$ and $s^{(1)}(T)$, $s^{(2)}(T)$, $s^{(3)}(T)$,
this way illustrating that the jump of the internal energy and the entropy at $T=T_p$ is conditioned by the terms with $m$ (thin red dotted curves).

\begin{figure}
\begin{center}
\includegraphics[clip=true,width=0.95\columnwidth]{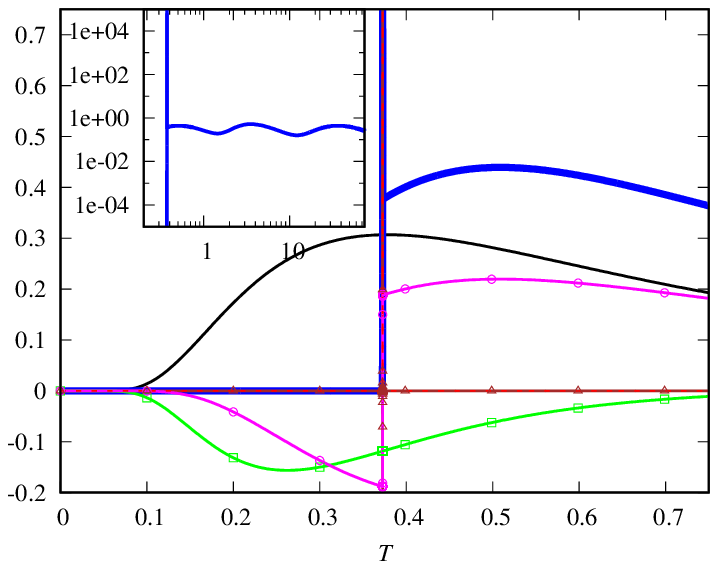}\\
\vspace{1mm}
\includegraphics[clip=true,width=0.95\columnwidth]{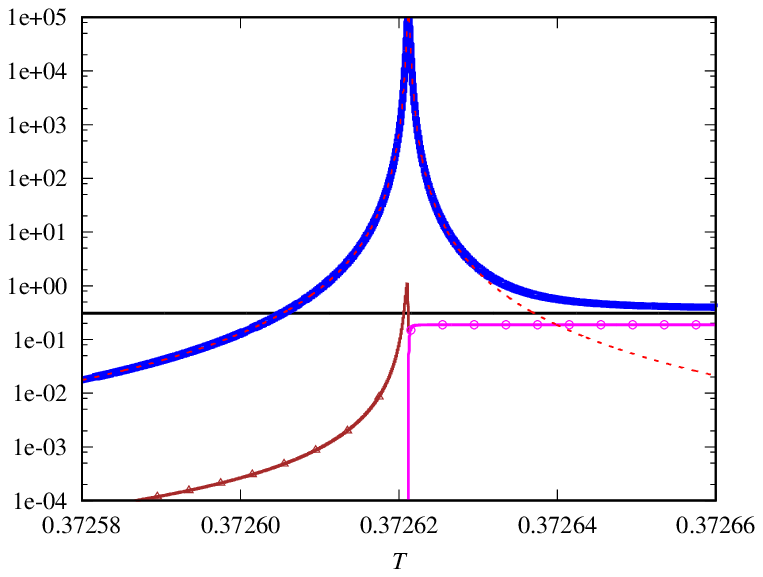}
\caption{Specific heat $c$ versus $T$; 
the spin-1/2 Ising-XYZ diamond chain (with $\gamma=0.7$, $h=12.7$) is described in Appendix~A, see Fig.~\ref{fig13}.
Thick blue curve corresponds to $c(T)$,
thin black, brown with triangles, green with squares, magenta with circles, and red dashed curves 
correspond to 
$c^{(1)}(T)$, $c^{(3)}(T)$, $c^{(4)}(T)$, $c^{(5)}(T)$, and $c^{(8)}(T)$, respectively,
see Eq.~(\ref{307}).}
\label{fig09}
\end{center}
\end{figure}

\begin{figure}
\begin{center}
\includegraphics[clip=true,width=0.95\columnwidth]{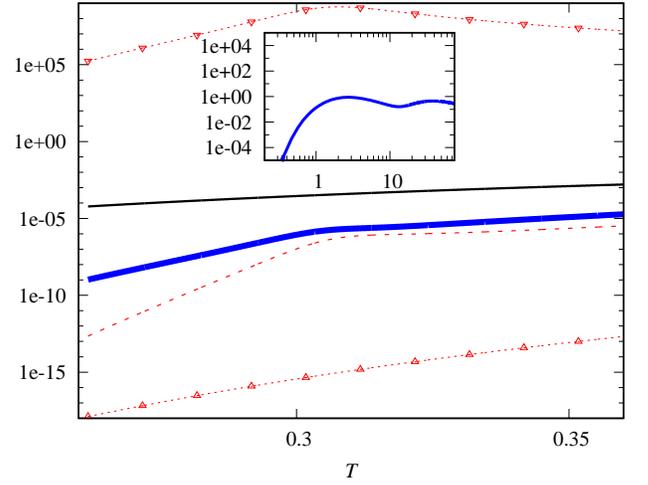}
\caption{Specific heat $c$ versus $T$; 
the spin-1/2 Ising-XYZ diamond chain (with $\gamma=0.82$, $h=7.916\,228$) is described in Appendix~A, see Fig.~\ref{fig13}.
Thick blue curve corresponds to $c(T)$,
thin black and red dashed curves 
correspond to 
$c^{(1)}(T)$ and $c^{(8)}(T)$, respectively,
see Eq.~(\ref{307}).
Moreover, red dashed curve with up-triangles and red dashed curve with down-triangles correspond to $(H^{\prime}_{\rm eff})^2$ and $T\chi$,
respectively,
see the formula for $c^{(8)}(T)$ in Eq.~(\ref{307}).}
\label{fig10}
\end{center}
\end{figure}

Finally,
we turn to the specific heat.
According to Eqs.~(\ref{307}) and (\ref{315}), (\ref{316}),
the dominant contribution to the specific heat around $T_p$ is expected to come from the susceptibility,
$c(T) \approx c^{(8)}(T)= T\chi(H_{\rm eff}^\prime)^2$,
and hence the specific heat manifests the behavior of the susceptibility.
More precisely, 
the temperature dependence of $c$ and $\chi$ around $T_p$ may be very similar,
since for $H_{\rm eff}\propto T-T_p$ the factor $T(H_{\rm eff}^\prime)^2$ around $T_p$ is only some finite constant,
see Fig.~\ref{fig09}.
However, it may happen that $T(H_{\rm eff}^\prime)^2$ is extremely small resulting in no peculiarity of the specific heat at $T_p$,
although such peculiarity does exist for the susceptibility,
see Fig.~\ref{fig10}.
Obviously,
to find the precise value of $c$ around $T_p$ we have to take into account all terms in Eq.~(\ref{307}).

In Fig.~\ref{fig09} we show $c(T)$ for $\gamma=0.7$ and $h=12.7$.
Of course, 
the result for the decorated spin chain coincides with the result for the effective Ising-chain model (\ref{201})
with the specific values of $C(T)$, $J_{\rm eff}(T)$, and $H_{\rm eff}(T)$
given in Eqs.~(\ref{a02}) and (\ref{a03}).
We can estimate different contributions $c^{(j)}(T)$, $j=1,\ldots,8$ in Eq.~(\ref{307}).
Only three terms (of eight) are relevant in the temperature region shown in Fig.~\ref{fig09}. 
The first term $c^{(1)}(T)$ (about $10^{-1}$) is conditioned by $C(T)$, since the contribution of the second term in Eq.~(\ref{301}) is less than $10^{-4}$.
The second term in Eq.~(\ref{307}), $c^{(2)}(T)$, is less than $10^{-5}$.
The term $c^{(3)}(T)=2T(\partial m/\partial T)H_{\rm eff}^{\prime}$ although is everywhere small (about $10^{-5}$) has the finite jump between almost $\pm 1$ at $T_p$.
The fourth term is a smooth function of $T$ having values about $10^{-1}$,
whereas the fifth one (about $10^{-1}$) contains $m$ and therefore has a finite jump at $T_p$.
Next two terms are about $10^{-7}$ and $10^{-6}$.
The most important at $T=T_p$ is the last (eighth) term in Eq.~(\ref{307}) which achieves the values about $10^{5}$.
However, outside a small vicinity of $T_p$, $c^{(8)}(T)$ is extremely small, see Fig.~\ref{fig06}.

It is worth noting 
that the height and width of the specific-heat peak at $T_p$ should not violate the thermodynamic relation $\int_0^\infty{\rm d}T c/T=\ln 2$,
i.e., a higher the peak is, a narrower it should be.

As it was mentioned above,
the relation between the specific heat and the susceptibility may be covered because of a small value of the factor $T(H_{\rm eff}^\prime)^2$ at $T_p$.
To illustrate such a case,
we consider the set of parameters which implies a small slope of the trajectory at the point where it crosses the vertical line ${\sf H}=0$ in Fig.~\ref{fig01}.
The results are reported in Fig.~\ref{fig10}.
The contribution of $c^{(8)}(T)$ (thin red dashed curve) to $c(T)$ (thick blue curve) does not yield any enhancement at $T=T_p$.
The reason for that becomes clear after inspecting the values of two factors of which $c^{(8)}(T)$ is consisted of,
that is,
$T\chi$ (red dashed with down-triangles curve)
and
$(H^{\prime}_{\rm eff})^2$ (red dashed with up-triangles curve).
While the first factor is about $10^{8}$ at $T=T_p\approx 0.312$ 
(that corresponds to ${\sf T}/{\sf J}\approx 0.099\,9$),
the second one is about $10^{-15}$ at $T=T_p$
resulting in no enhancement of the specific heat at $T=T_p$.

To summarize this section,
we have demonstrated that the low-temperature peculiarities of the effective Ising-chain model 
(and thus of the decorated spin chains) 
are related to the critical point of the Ising-chain model ${\sf H}=0$ and ${\sf T}=0$.
This is obvious from consideration of the correlation length $\xi(T)$ in the ${\sf H}/{\sf J}-{\sf T}/{\sf J}$ plane,
Fig.~\ref{fig04}.
Both, the magnetization and the susceptibility straightforwardly reflect the low-temperature behavior of the ferromagnetic Ising-chain model,
Figs.~\ref{fig05} and \ref{fig06}.
The magnetization $m(T)$ has almost not-smeared jump between the two saturation values of opposite signs at $T=T_p$
and 
the susceptibility $\chi(T)$ exhibits an abrupt increase at $T=T_p$ which reaches the value (\ref{315})
(which corresponds to the zero-field low-temperature value of $\chi$ for the ferromagnetic Ising-chain model).
The internal energy and the entropy for the effective Ising-chain model with temperature-dependent parameters (\ref{201})
depend on the magnetization, see Eqs.~(\ref{305}) and (\ref{306}),
and therefore the magnetization jump at $T_p$ shows up in the temperature profiles of $e$ and $s$, too.
Moreover, in contrast to the standard ferromagnetic Ising-chain model, 
the specific heat for the effective Ising-chain model (\ref{201}) exhibits a sharp maximum at $T_p$;
according to Eq.~(\ref{307}),
this can be traced to the abrupt increase of the susceptibility at $T=T_p$
if it is not quenched by the factor $T(H_{\rm eff}^\prime)^2$ at $T=T_p$.

It should be also noted that the temperatures at which 
the effective field $H_{\rm eff}$ vanishes ($T_p$), 
the correlation length has a peak ($T_{\rm max.cor.l.}$), 
the susceptibility has a peak ($T_{\rm max.susc.}$), 
or 
the specific heat has a peak ($T_{\rm max.sp.heat}$)
are, generally speaking, not identical.
However,
for the case $\gamma=0.7$, $h=12.7$ 
we have $T_p=0.372\,621\,188\,80$ and these 11 digits for all characteristic temperatures coincide.
(For the case $\gamma=0.7$, $h=12$ 
we have $T_p=0.815\,048\,150\,05$ and only first 4 digits for all characteristic temperatures coincide.)

\section{Universality}
\label{sec4}
\setcounter{equation}{0}

As it immediately follows from explanations of the previous section, 
the pseudo-critical behavior is universal and depends 
1) on the fact that the Hamiltonian parameters are temperature dependent,
Eqs.~(\ref{305}), (\ref{306}), and (\ref{307}); 
2) on the critical behavior of the standard Ising-chain model around its critical point ${\sf H}=0$ and ${\sf T}=0$; 
and also
3) on the specific temperature dependence of the Hamiltonian parameters
[especially of $H_{\rm eff}(T)$]
around $T=T_p$.

Thus,
Eqs.~(\ref{314}) and \eqref{316} say that as $T$ approaches $T_p$,
$\xi(T)\propto \vert H_{\rm eff}(T)\vert^{-1}$
and
$\chi(T)\propto \vert H_{\rm eff}(T)\vert^{-3}$.
Moreover,
Eq.~(\ref{307}), when $c^{(8)}(T)$ is relevant, suggests that as $T$ approaches $T_p$,
$c(T)\propto \chi(T)\propto \vert H_{\rm eff}(T)\vert^{-3}$.
Therefore, the relations $\alpha=\alpha^\prime =\gamma=\gamma^\prime =3\nu=3\nu^\prime$ are obvious.
Further on, since for the decorated spin chains at hand
\begin{eqnarray}
\label{401}
H_{\rm eff}(T)
\xrightarrow[ T\to T_p ]{}
A\left(T-T_p\right),
\end{eqnarray}
we immediately obtain 
$\alpha=\alpha^\prime=3$,
$\gamma=\gamma^\prime=3$, 
and
$\nu=\nu^\prime=1$ \cite{Rojas2018b}.

In the vicinity of $T_p$ it would be sufficient to consider $J_{\rm eff}(T)=J_{\rm eff}>0$ and $H_{\rm eff}(T)=A\left(T-T_p\right)$ only.
But these parameters cannot reproduce the whole range of temperatures 
leading to such shortcomings as negative entropy, specific heat etc. outside the vicinity of $T_p$.

We have also to underline the role of the slope related to the factor $A$ in Eq.~\eqref{401}.
As we have demonstrated in Sec.~\ref{sec3}
(see discussion around Fig.~\ref{fig10}),
a small value of $T(H_{\rm eff}^\prime)^2\vert_{T=T_p}=T_pA^2$ may quench the peculiarity coming from $\chi(T\to T_p)$.

\section{Conclusions}
\label{sec5}
\setcounter{equation}{0}

\begin{figure}
\begin{center}
\includegraphics[clip=true,width=0.95\columnwidth]{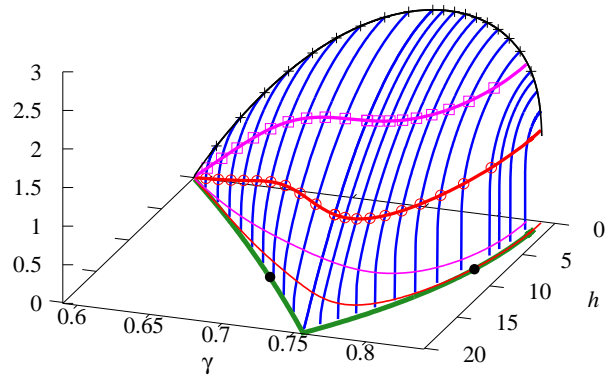}
\caption{Phase diagram: $T_p$ in the $\gamma-h$ plane; 
the spin-1/2 Ising-XYZ diamond chain is described in Appendix~A, see Fig.~\ref{fig13}.
Blue curves correspond to $T_p(\gamma,h)$;
thick red curve with empty circles denotes $T_p$ at which $\xi(\gamma,h,T_p)=100$ and $2J_{\rm eff}(T_p)/T_p\approx 5.298$
(thin red curve is the projection of the thick red curve on the $\gamma-h$ plane);
thick magenta curve with empty squares denotes $T_p$ at which $\xi(\gamma,h,T_p)=10$  and $2J_{\rm eff}(T_p)/T_p\approx 2.997$
(thin magenta curve is the projection of the thick magenta curve on the $\gamma-h$ plane);
forest-green curve denotes the region with $H_{\rm eff}<0$ at $T=0$ (cf. Fig.~\ref{fig13}).
Two black circles denote the points $\gamma=0.7$, $h=12.7$ and $\gamma=0.82$, $h=7.916\,228$.}
\label{fig11}
\end{center}
\end{figure}

Let us summarize the present study.
We have not reported many new calculations, 
rather we have suggested a new perspective for the temperature-driven pseudo-transitions in one-dimensional decorated spin systems with short-range interactions.
First of all,
we have illustrated usefulness of the mapping onto the standard Ising-chain model. 
We have stressed that the observed low-temperature peculiarities of the decorated spin chains are related 
1) to the critical point of the Ising-chain model ${\sf H}=0$ and ${\sf T}=0$ 
and 
2) to the specific temperature dependences of the effective parameters of the effective Ising-chain model which represents the initial decorated spin chain.
We have further discussed the necessary \cite{Souza2018,Carvalho2018} and sufficient conditions for occurrence of the pseudo-transition:
While the necessary condition (\ref{202}) requires $H_{\rm eff}=0$ (and hence ${\sf H}=0$) at $T=T_p$,
the sufficient condition (\ref{310}) says that ${\sf T}/{\sf J}\ll 2$ at $T=T_p$
(i.e., the temperature for the standard Ising-chain model without field should be sufficiently low).
In Fig.~\ref{fig11} we illustrate these arguments for the spin-1/2 Ising-XYZ diamond chain with $J=100$, $J_z=24$, and $J_0=-24$ in the $\gamma-h$ plane:
All points in a triangle which is singled out by the forest-green curve and the straight line $h=0$ 
satisfy the necessary condition (\ref{202}) for the existence of the pseudo-transition.
The values of $T_p$ are given by the blue lines.
However, 
only sufficiently close to the forest-green line one can satisfy the sufficient condition (\ref{310}),
i.e., one can observe a developing of the sufficiently large correlation length (\ref{309})
which causes peculiarities in the low-temperature properties of the decorated spin chains.
Nonetheless, even in this case, just the specific heat may show no enhancement at $T_p$ as was illustrated in Fig.~\ref{fig10}.
Finally,
we have explained the power-law behavior of various quantities in the vicinity of the pseudo-critical temperature.

We think, that the elaborated perspective has several further extensions which deserve to be studied.
The most straightforward one is related to the decorated spin chains which can be reduced to the effective higher-spin Ising-chain models.
Moreover,
we believe that the decorated spin models are of some interest not only in one dimension
and the case of more than one dimensions may be also intriguing.
Although in two dimensions there is the famous Ising-Onsager transition,
a two-dimensional decorated model which can be reduced to the square-lattice Ising model 
with particular trajectories around the critical point ${\sf H}=0$ and ${\sf T}=2{\sf J}/\ln(1+\sqrt{2})$ in the ${\sf H}/{\sf J}-{\sf T}/{\sf J}$ plane
should also exhibit interesting behavior conditioned by that critical point. 
As a candidate for such a decorated two-dimensional spin model we may suggest a diamond-like-decorated square lattice \cite{Hirose2017}.

\section*{Acknowledgments} 

The authors gratefully acknowledge helpful discussions with 
Yu.~Holovatch,
A.~Honecker,
J.~Richter,
A.~Shvaika,
I.~V.~Stasyuk,
J.~Stre\v{c}ka, 
and 
T.~Verkholyak.
O.~D. was supported by the Brazilian agency FAPEMIG (CEX - BPV-00090-17); 
he appreciates the kind hospitality of the Federal University of Lavras in October-December of 2017 when this study was launched.
O.~D. acknowledges the kind hospitality of J.~Stre\v{c}ka (Pavol Jozef \v{S}af\'{a}rik University in Ko\v{s}ice, Slovakia)
during 
the 17th Czech and Slovak Conference on Magnetism (June 3-7, 2019)
and
the Workshop on Quantum Magnetism: Theoretical Challenges and Future Perspectives (June 7-8, 2019).

\section*{Appendix A: Spin-1/2 Ising-XYZ diamond chain}
\renewcommand{\theequation}{A.\arabic{equation}}
\setcounter{equation}{0}

\begin{figure}
\begin{center}
\includegraphics[clip=true,width=0.95\columnwidth]{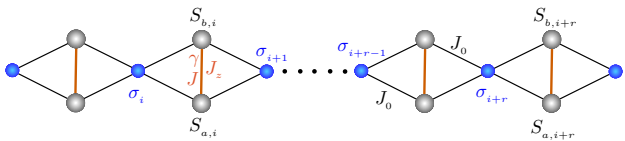}
\caption{Spin-1/2 Ising-XYZ diamond chain lattice;
$\sigma_i$ denotes the Ising spin 1/2 and $S_{a,i}$ and $S_{b,i}$ correspond to the Heisenberg spins 1/2.
For further details see Ref.~\cite{Carvalho2019}.}
\label{fig12}
\end{center}
\end{figure}

\begin{figure}
\begin{center}
\includegraphics[clip=true,width=0.95\columnwidth]{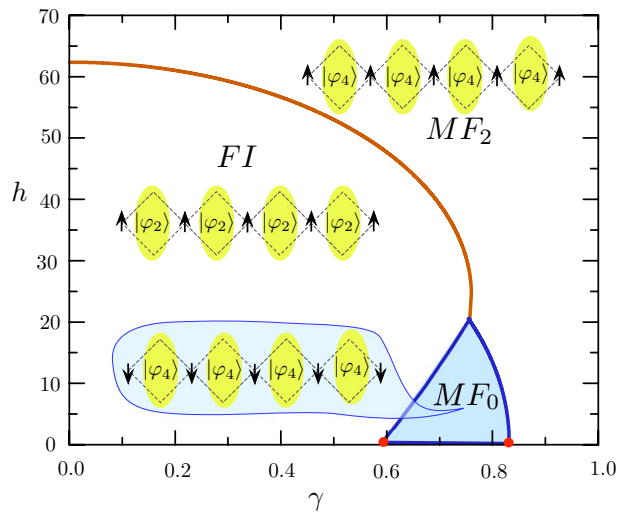}
\caption{Ground-state phase diagram for the spin-1/2 Ising-XYZ diamond chain (\ref{a01}) with $J=100$, $J_z=24$, and $J_0=-24$
in the $\gamma-h$ plane.
Two red points correspond to $\gamma\approx 0.5892$ and $\gamma\approx 0.8314$.
For further details see Ref.~\cite{Carvalho2019}.}
\label{fig13}
\end{center}
\end{figure}

We consider the spin-1/2 Ising-XYZ diamond chain with the Hamiltonian
(we use notations of Ref.~\cite{Carvalho2019}):
\begin{eqnarray}
\label{a01}
{\cal H}=\sum_{i=1}^{\sf N} H_i,
\nonumber\\
H_i=-J(1\!+\!\gamma)S_{a,i}^xS_{b,i}^x -J(1\!-\!\gamma)S_{a,i}^yS_{b,i}^y -J_zS_{a,i}^zS_{b,i}^z
\nonumber\\
-J_0(S_{a,i}^z+S_{b,i}^z)(\sigma_i+\sigma_{i+1})
\nonumber\\
-h_z(S_{a,i}^z+S_{b,i}^z)-\frac{h}{2}(\sigma_i+\sigma_{i+1}),
\end{eqnarray}
see Fig.~\ref{fig12}.
Here $S^\alpha_{a(b)}$ ($\alpha = x, y, z$) are the spin-1/2 operators, 
$\sigma$ corresponds to the Ising spins 1/2 , 
$\gamma$ is the XY-anisotropy parameter, 
$J$ and $J_z$ are the Heisenberg-like interactions between interstitial sites, 
the exchange parameter $J_0$ represents the Ising-like interaction between nodal and interstitial sites, 
and the external magnetic fields $h_z$ and $h$ are assumed to be along the $z$ direction. 
Later on,
it is assumed that $h_z=h\ge 0$.
Note also that the Ising spin 1/2 in Eq.~(\ref{a01}) is two times smaller than the Ising spin in Eq.~\eqref{201}.

The ground-state phase diagram in the $\gamma-h$ plane for a particular set of coupling parameters
$J=100$, $J_z=24$, and $J_0=-24$ is shown in Fig.~\ref{fig13}.
It contains three phases:
One ferrimagnetic phase ($FI$)
and
two modulated ferromagnetic Heisenberg phases ($MF_0$ and $MF_2$),
see Ref.~\cite{Carvalho2019} and references therein.

The spin-1/2 Ising-XYZ diamond chain model (\ref{a01}) can be mapped onto the standard Ising-chain model 
with the Hamiltonian given in Eq.~\eqref{201}
and
\begin{eqnarray}
\label{a02}
C=-{\sf N}\frac{T}{4}\ln\left(w_1w^2_0w_{-1}\right),
\nonumber\\
J_{\rm eff}=\frac{T}{4}\ln\frac{w_1 w_{-1}}{w_{0}^2},
\nonumber\\
H_{\rm eff}=\frac{T}{2}\ln\frac{w_1}{w_{-1}}.
\end{eqnarray}
Here
\begin{eqnarray}
\label{a03}
w_\mu=2\exp\frac{h\mu}{2T}
\nonumber\\
\times
\left[
\exp\left(-\frac{J_z}{4T}\right)\cosh\frac{J}{2T}+\exp\frac{J_z}{4T}\cosh\frac{\Delta_\mu}{T}
\right],
\nonumber\\
\Delta_\mu=\sqrt{\left(h_z+J_0\mu\right)^2+\frac{\gamma^2J^2}{4}}.
\end{eqnarray}
Importantly,
$J_{\rm eff}> 0$ at all temperatures $T$ for all values of $\gamma$ and $h$ in Fig.~\ref{fig13}.
In contrast, 
$H_{\rm eff}> 0$ at all temperatures $T$ for all values of $\gamma$ and $h$ in Fig.~\ref{fig13} except those, which belong to the $MF_0$ phase.
For the values of $\gamma$ and $h$ which belong to the $MF_0$ phase,
$H_{\rm eff}< 0$ at $T=0$ but becomes positive for high temperatures,
see, e.g., Fig.~\ref{fig01}.

\section*{Appendix B: Coupled spin-electron double-tetrahedral chain}
\renewcommand{\theequation}{B.\arabic{equation}}
\setcounter{equation}{0}

\begin{figure}
\begin{center}
\includegraphics[clip=true,width=0.95\columnwidth]{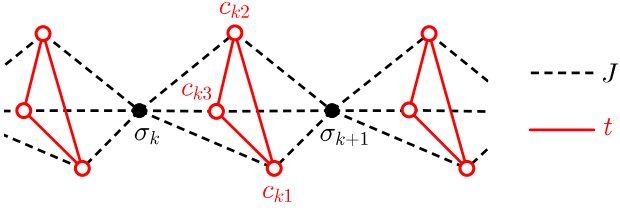}
\caption{A part of the spin-electron system on a double-tetrahedral chain. 
Full circles denote nodal lattice sites occupied by the localized Ising spins 1/2, 
while the empty circles forming triangular plaquettes are available to mobile electrons
(each three equivalent sites of the triangular plaquette are available to two mobile electrons).
For further details see Ref.~\cite{Galisova2015}.}
\label{fig14}
\end{center}
\end{figure}

\begin{figure}
\begin{center}
\includegraphics[clip=true,width=0.95\columnwidth]{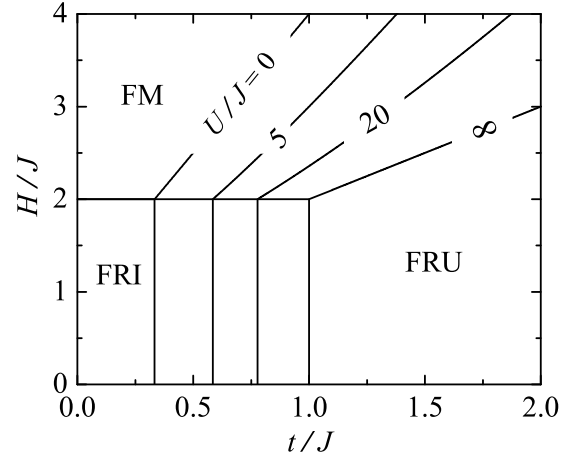}
\caption{The ground-state phase diagram in the $t/J-H/J$ plane 
for the coupled spin-electron double-tetrahedral chain (\ref{b01}) with the antiferromagnetic coupling $J >0$
upon varying a relative strength of the Coulomb term $U/J= 0,\,5,\,20,\,\infty$.
For further details see Ref.~\cite{Galisova2015}.}
\label{fig15}
\end{center}
\end{figure}

We consider the double-tetrahedral chain (Fig.~\ref{fig14}) 
of localized Ising spins 1/2 and mobile electrons (two mobile electrons on each triangular plaquette)
with the Hamiltonian
(we use notations of Ref.~\cite{Galisova2015}):
\begin{eqnarray}
\label{b01}
{\cal H}=\sum_{i=1}^{\sf N} H_i,
\nonumber\\
H_i=\!-t\!\!\!\sum_{\alpha=\uparrow,\downarrow}\!\!\!
\left(c^\dagger_{i1,\alpha}c_{i2,\alpha}\! +\! c^\dagger_{i2,\alpha}c_{i3,\alpha} \!+\! c^\dagger_{i3,\alpha}c_{i1,\alpha}\! +\!{\rm H.c.} \right)
\nonumber\\
+U\sum_{j=1}^3n_{ij,\uparrow}n_{ij,\downarrow}
+\frac{J}{2}\left(\sigma_i^z+\sigma_{i+1}^z\right)\sum_{j=1}^3\left(n_{ij,\uparrow}-n_{ij,\downarrow}\right)
\nonumber\\
-\frac{H_{I}}{2}\left(\sigma_i^z+\sigma_{i+1}^z\right)-\frac{H_e}{2}\sum_{j=1}^3\left(n_{ij,\uparrow}-n_{ij,\downarrow}\right),
\end{eqnarray}
see Fig.~\ref{fig14}.
Above, 
$c_{ij,\alpha}^{\dagger}$ and $c_{ij,\alpha}$ represent usual fermionic creation and annihilation operators for mobile electrons 
from the $i$th triangular plaquette 
with spin $\alpha = \uparrow$ or $\downarrow$, 
$n_{ij,\alpha}=c_{ij,\alpha}^{\dagger}c_{ij,\alpha}$ is the respective number operator, 
$\sigma_i^z = \pm 1/2$ labels the Ising spin placed at the $i$th nodal lattice site, 
and $\sf N$ denotes the total number of nodal lattice sites. 
The hopping parameter $t > 0$ takes into account the kinetic energy of mobile electrons delocalized over triangular plaquettes, 
$U \ge 0$ represents the on-site Coulomb repulsion between two electrons of opposite spins occupying the same lattice site, 
and $J$ stands for the Ising coupling between the mobile electrons and their nearest Ising-spin neighbors. 
Finally, the fields $H_I$ and $H_e$ enter the Zeeman's terms 
accounting for the magnetostatic energy of the localized Ising spins and mobile electrons in the presence of an external magnetic field.
Later on, it is assumed that $H_I=H_e\equiv H\ge 0$.
Note also that the Ising spin 1/2 in Eq.~(\ref{b01}) is two times smaller than the Ising spin in Eq.~\eqref{201}.

The ground-state phase diagram in the $t/J-H/J$ plane for $J>0$ and several values of $U/J=0,\,5,\,20,\,\infty$ is shown in Fig.~\ref{fig15}.
It contains three different states:
The ferromagnetic (FM) state,
the ferrimagnetic (FRI) state,
and
the frustrated (FRU) state.

The model \eqref{b01} can be mapped onto the standard Ising-chain model 
with the Hamiltonian given in Eq.~\eqref{201}
and
\begin{eqnarray}
\label{b02}
C=-{\sf N}\frac{T}{4}\ln\left[(W_-+W)(W_++W)(W_0+W)^2\right],
\nonumber\\
J_{\rm eff}=\frac{T}{4}\ln\frac{(W_-+W)(W_++W)}{(W_{0}+W)^2},
\nonumber\\
H_{\rm eff}=\frac{H_I}{2}+\frac{T}{2}\ln\frac{W_-+W}{W_{+}+W}.
\end{eqnarray}
Here
\begin{eqnarray}
\label{b03}
W_{\mp}
\!=\!
\left[2\exp\frac{t}{T}+\exp\left(\!\!-\frac{2t}{T}\right)\right]\!\!
\left(1+2\cosh\frac{J\mp H_e}{T}\!\right),
\nonumber\\
W_0
\!=\!
\left[2\exp\frac{t}{T}+\exp\left(\!\!-\frac{2t}{T}\right)\right]\!\!
\left(1+2\cosh\frac{H_e}{T}\right),
\nonumber\\
W
=
4\exp\left(-\frac{t+U}{2T}\right)\cosh\frac{\sqrt{(U-t)^2+8t^2}}{2T}
\nonumber\\
+2\exp\frac{2t-U}{2T}\cosh\frac{\sqrt{(U+2t)^2+32t^2}}{2T}.
\end{eqnarray}
The authors of Ref.~\cite{Galisova2015} discovered unexpected low-temperature behavior 
and considered in some detail as an example the set of parameters
$U/J=5$, $t/J=0.6$, $H/J=0,\ldots,2$, and $T/J=0,\ldots,10$.
(Figs.~6 and 7 of Ref.~\cite{Galisova2015}).
For this set of parameters $J_{\rm eff}> 0$ at all temperatures,
whereas $H_{\rm eff}$ may change its sign twice, see Fig.~\ref{fig02}.
Only at the smaller temperature $T_p$ which yields $H_{\rm eff}(T_p)=0$ and satisfies (\ref{310})
the peculiarities in thermodynamic quantities are clearly seen.

\section*{Appendix C: Spin-1/2 Ising-Heisenberg double-tetrahedral chain}
\renewcommand{\theequation}{C.\arabic{equation}}
\setcounter{equation}{0}

\begin{figure}
\begin{centering}
\includegraphics[scale=0.6]{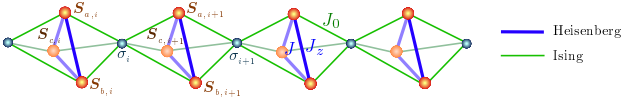}
\par
\end{centering}
\caption{Schematic representation of the Ising-Heisenberg  double-tetrahedral chain. 
Small balls ($\sigma_{i}$) correspond to Ising spins 
and 
large balls ($\boldsymbol{S}_{a(b),i}$) correspond to Heisenberg spins.
For further details see Ref.~\cite{Rojas2018a}.}
\label{fig16}
\end{figure}

\begin{figure}
\begin{centering}
\includegraphics[scale=0.35]{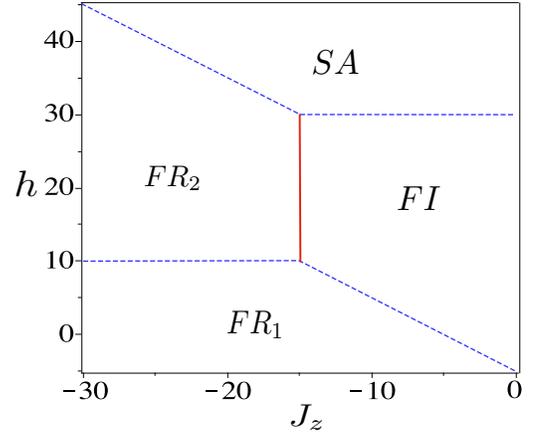}
\par
\end{centering}
\caption{Zero-temperature phase diagram in the $J_{z}-h$ plane, 
assuming fixed parameters $J=-10$, $J_{0}=-10$, and $h_{z}=h$.
For further details see Ref.~\cite{Rojas2018a}.}
\label{fig17}
\end{figure}

Previously, in Refs.~\cite{Mambrini1999,Rojas2003,Maksymenko2011} the pure Heisenberg double-tetrahedral chain was considered.
Later, in Refs.~\cite{Ohanyan2009,Antonosyan2009} the Ising-Heisenberg version of the model was introduced (see Fig. \ref{fig16}).
Although the latter model was discussed in Refs.~\cite{Ohanyan2009,Antonosyan2009}, 
the pseudo-transition property has been explored only recently in Ref.~\cite{Rojas2018a}.
The corresponding Hamiltonian of this model is 
\begin{eqnarray}
\label{c01}
H=-\sum_{i=1}^{\sf N}\left\{ J(\boldsymbol{S}_{a,i},\boldsymbol{S}_{b,i})_{z}+J(\boldsymbol{S}_{b,i},\boldsymbol{S}_{c,i})_{z}\right.
\nonumber\\
+J(\boldsymbol{S}_{c,i},\boldsymbol{S}_{a,i})_{z}+\tfrac{h}{2}\left(\sigma_{i}+\sigma_{i+1}\right)
\nonumber\\
\left.
+\left(S_{a,i}^{z}+S_{b,i}^{z}+S_{c,i}^{z}\right)\left[h_{z}+J_{0}(\sigma_{i}+\sigma_{i+1})\right]\right\},
\end{eqnarray}
where $J(\boldsymbol{S}_{a,i},\boldsymbol{S}_{b,i})_{z}\equiv JS_{a,i}^{x}S_{b,i}^{x}+JS_{a,i}^{y}S_{b,i}^{y}+J_{z}S_{a,i}^{z}S_{b,i}^{z}$
with $S_{a,i}^{\alpha}$ denoting the Heisenberg spin-1/2 and $\alpha=\{x,y,z\}$,
while $\sigma_{i}$ denotes the Ising spin ($\sigma_{i}=\pm 1/2$).
In a similar way the Heisenberg operators are defined for sites $b$ and $c$ in \eqref{c01}.

The zero-temperature phase diagram of this model is reported in Fig.~\ref{fig17}. 
We observe there four different states, for details see Ref.~\cite{Rojas2018a}. 
Here we illustrate the phase boundary between the frustrated phase ($FR_{2}$) and the ferrimagnetic phase ($FI$) by a solid line, 
where the pseudo-transition shows up in the low-temperature region.

For the present model, the Boltzmann factor was obtained in Ref.~\cite{Rojas2018a};
it can be expressed as follows: 
\begin{eqnarray}
\label{c02}
w_{n}
=
2\exp\frac{2hn-J_z}{4T}
\nonumber\\
\times
\left\{
\left[\exp\frac{J}{T}+2\exp\left(-\frac{J}{2T}\right)\right]\cosh\frac{h_z+J_0n}{2T}
\right. 
\nonumber\\
\left.
+\exp\frac{J_z}{T}\cosh\frac{3\left(h_z+J_0n\right)}{2T}
\right\},
\end{eqnarray}
where $n=\{-1,0,1\}$.
The effective parameters in Eq.~(\ref{201}) are determined through $w_n$ (\ref{c02}) as follows:
\begin{eqnarray}
\label{c03}
C=-\mathsf{N}\frac{T}{4}\ln\left(w_{1}w_{0}^{2}w_{-1}\right),
\nonumber\\
J_{\mathrm{eff}}= \frac{T}{4}\ln\frac{w_{1}w_{-1}}{w_{0}^{2}},
\nonumber\\
H_{\mathrm{eff}}= \frac{T}{2}\ln\frac{w_{1}}{w_{-1}}.
\end{eqnarray}

\end{document}